\documentclass[11pt]{article}
\usepackage{amssymb}
\usepackage{latexsym}
\begin{document}
\setlength{\baselineskip}{15pt}
\title{Poisson systems as the natural framework for \\
additional first integrals via Darboux invariant hypersurfaces $^{\star}$}
\author{Isaac A. Garc\'{\i}a$^{a}$, \mbox{}  Benito Hern\'{a}ndez--Bermejo$^{b,*}$}
\date{}

\maketitle

\begin{description}
\item[$^a$] {\em Departament de Matem\`atica, Universitat de Lleida, Avda. Jaume II, 69, 25001 Lleida, Spain.}
\item[$^b$] {\em Departamento de F\'{\i}sica. Universidad Rey Juan Carlos. Calle Tulip\'{a}n S/N. 28933 M\'{o}stoles, Madrid. Spain.}
\end{description}

\mbox{}

\begin{center}
{\bf Abstract}
\end{center}
\noindent
In the literature, the existence of Darboux polynomials and additional polynomial first integrals has been considered in the case of Hamiltonian systems. In this article such problem is formulated in the more general framework of Poisson structures, which include Hamiltonian systems as a particular case. This generalization allows a natural extension of the previous results, which can now be applied to a larger class of vector fields and is valid for arbitrary diffeomorphisms (instead of canonical transformations). Examples are discussed.

\mbox{}

\mbox{}

\mbox{}

\mbox{}

\noindent {\bf AMS(2010):} 34C20, 37J35. 

% 34C20 Transformation and reduction of equations and systems, normal forms. 
% 37J35 Completely integrable systems, topological structure of phase space, 
%          integration methods. 

\mbox{}


% 45.20.-d Formalisms in classical mechanics.
% 45.20.Jj Lagrangian and Hamiltonian mechanics.

\mbox{}


\vfill

\footnoterule

\noindent $^\star$ The first author is partially supported by a MCYT/FEDER grant number MTM2008-00694 and by a CIRIT grant number 2009 SGR 381.

\noindent $^*$ Corresponding author. Telephone: (+34) 91 488 73 91. Fax: (+34) 91 664 74 55.

\noindent {\em E-mail addresses:} garcia@matematica.udl.cat (I.A. Garc\'{\i}a),
benito.hernandez@urjc.es (B. Hern\'{a}ndez-Bermejo).

\pagebreak

\begin{flushleft}
{\bf 1. Introduction and background}
\end{flushleft}

Hamiltonian systems have a paramount significance in physics from a fundamental point of view as well as from an applied perspective. In particular, some recent contributions \cite{MP,GGL} have analyzed the role of Darboux polynomials in the integrability of natural polynomial Hamiltonian systems.

Consider a differential autonomous system $\dot{x} = f(x)$ where $f : \Omega \subseteq \mathbb{R}^{n} \to \mathbb{R}^{n}$ is of class $C^1(\Omega)$. In components we have $x =(x_1, \ldots, x_n)$ and $f(x) =(f_1(x), \ldots, f_n(x))$. We denote by $\mathcal{X}$ the associated vector field $\mathcal{X} = \sum_{i=1}^n f_i(x) \partial_{x_i}$. We say that $\mathcal{X}$ has a (proper) {\it Darboux invariant hypersurface} $\Sigma$ with respect to $\Phi$ if $\Sigma \subset \Omega$ is an invariant set under the flow of $\mathcal{X}$ defined by $\Sigma = \{ x \in \Omega : F(x) = 0 \}$ with $F : \Omega \to \mathbb{R}$ non-constant and $F \in C^1(\Omega)$ such that $F(x)$ is not a first integral of $\mathcal{X}$ and there is a $C^1(\Omega)$-map $\Phi : \Omega \to \mathbb{R}^n$ satisfying that $F \circ \Phi$ is a polynomial in $\mathbb{R}[\Phi]$.

This definition is a generalization of the classical (proper) Darboux polynomial first defined and used for integrability purposes in \cite{Da}. Recall that when $\mathcal{X}$ is a polynomial vector field and $\Phi$ is the identity map, then $F$ becomes a non-constant polynomial $F \in \mathbb{R}[x]$ and is called (proper) {\em Darboux polynomial}. Of course, in this case $\Sigma$ is an algebraic invariant hypersurface of $\mathcal{X}$ and there exists a polynomial $K \in \mathbb{R}[x]$ called the {\em cofactor} of $F$ such that $\mathcal{X} F =  K F$. Notice that, since $F$ is proper, its cofactor $K$ is not identically zero. The Darboux theory of integrability has been strongly developed in the past years, see for instance 
\cite{Llibre-book} and references therein.

As indicated, the physical significance of Hamiltonian systems is difficult to overemphasize.  However, Hamiltonian theory (and in particular the inherent limitation derived from the use of canonical transformations) defines in practice an operationally restrictive context, as we shall see. The main goal of this work is to show that the framework provided by Poisson systems is actually the most natural for the analysis of the results developed in \cite{MP,GGL} leading, in fact, to a significant generalization of them.

Finite-dimensional Poisson systems are well-known in most fields of physics both for their formal interest and for providing the basis for many different analytical and numerical tools of applied interest (e.g. see the discussions in \cite{olv1}-\cite{GB} for a brief account of these methods and application domains). When expressed in coordinates $x_1, \ldots , x_n$, a smooth dynamical system defined in a domain $\Omega \subset \mathbb{R}^n$ is said to be a finite-dimensional Poisson system if it can be written in the form
\begin{equation}
    \label{poiss}
    \dot{x} \equiv \frac{\mbox{d}x}{\mbox{d}t}= {\cal J}(x) \cdot \nabla H(x)
\end{equation}
where $x = (x_1, \ldots ,x_n)^T$, superscript $^T$ denotes the transpose of a matrix, function $H(x)$ is by construction a time-independent first integral (the Hamiltonian), and the $n \times n$ {\em structure matrix\/} ${\cal J}(x)$ is composed by the {\em structure functions\/} $J_{ij}(x)$ which must verify the Jacobi PDEs:
\[
     \sum_{l=1}^n
	\left( \begin{array}{c} J_{li}(x) \partial_l J_{jk}(x) + J_{lj}(x) \partial_l J_{ki}(x) +
     J_{lk}(x) \partial_l J_{ij}(x) \end{array} \right) = 0
	\:\; , \;\:\;\: i,j,k=1, \ldots ,n
\]
where $ \partial_l \equiv \partial / \partial x_l$. The structure functions must be also
skew-symmetric: $J_{ij}(x) =  - J_{ji}(x)$ for all $i,j=1, \ldots ,n$. The relevance of Poisson dynamical systems relies on several reasons. One is that they provide a broad generalization of classical Hamiltonian systems, allowing not only for odd-dimensional vector fields, but also because Poisson structure matrices admit a great diversity of forms apart from the fixed one associated to the Hamiltonian case. Actually, Poisson systems are a generalization of classical Hamiltonian systems on which a generalized Poisson bracket is defined. For what is to follow, an important feature of Poisson systems is that they are not restricted by the use of canonical transformations. In fact, every diffeomorphic change of variables $y_i = y_i(x)$, maps the Poisson system (\ref{poiss}) into another Poisson system of Hamiltonian $H^*(y)=H(x(y))$ and structure matrix ${\cal J}^*(y)$ of entries
\begin{equation}
\label{jdiff}
      J^*_{ij}(y) = \sum_{k,l=1}^n \frac{\partial y_i}{\partial x_k} J_{kl}(x)
	\frac{\partial y_j}{\partial x_l} \;\: , \;\:\;\: i,j = 1, \ldots ,n
\end{equation}
When Rank($\cal{J}$)$=n$ the Poisson system is termed {\em symplectic\/}. The possible rank degeneracy of the structure matrix ${\cal J}$ implies that a certain class of first integrals ($D(x)$ in what follows) termed {\em Casimir invariants\/} exist. There is no analog in the framework of classical Hamiltonian systems for such constants of motion, which are characterized as the solution set of the system of coupled PDEs: ${\cal J} \cdot \nabla D =0$. The determination of Casimir invariants and their use in order to carry out a reduction (local, in principle) is the cornerstone of the (at least local) equivalence between Poisson systems and classical Hamiltonian systems, as stated by Darboux theorem \cite{olv1}. This justifies that Poisson systems can be regarded, to a large extent, as a natural generalization of classical Hamiltonian systems. In this context, the following matrix conventions shall be used. In first place, the structure matrix of an $m$-degree of freedom classical Hamiltonian system will be termed
\[
	\mathcal{S}_{2m} = \left( \begin{array}{cc} O_m & I_m \\ -I_m & O_m \end{array} \right)
\]
with $I_m$ and $O_m$ the $m \times m$ identity and null matrices, respectively. 
Additionally, the structure matrix corresponding to the Darboux canonical form of a Poisson system of rank $2m$ having $s$ (decoupled) Casimir invariants will be:
\[
	\mathcal{S}_{2m,s} = \mathcal{S}_{2m} \oplus O_s =
	\left( \begin{array}{ccc} O_m & I_m & \mbox{} \\ -I_m & O_m & \mbox{} \\
	\mbox{} & \mbox{} & O_s \end{array} \right)
\]
A first integral $I(x)$ of the Poisson system with Hamiltonian $H(x)$ and a complete set of independent Casimir invariants $D_1(x), \ldots, D_s(x)$ will be called an {\em additional first integral} when $I$ is functionally independent of $H$ and all the $D_i$. Recall that two functions are functionally independent when the gradient vectors are linearly independent except perhaps in a zero Lebesgue measure set.

As indicated, the present work is a natural extension of the references \cite{MP,GGL} and our goal is twofold: in first place, to show that Poisson systems provide a natural framework for the investigation of first integrals derived from the existence of proper Darboux hypersurfaces; and second, to make use of such framework in order to generalize the results in \cite{MP,GGL}. The structure of the article is the following. In Section 2 it is shown how such results ca be generalized, still in a purely Hamiltonian context. Section 3 provides the extension to the Poisson case in the symplectic situation. The generalization to general (possibly non-symplectic) Poisson systems is the goal of Section 4.

%\mbox{}
\pagebreak
\begin{flushleft}
{\bf 2. Case I: Linear canonical transformation approach}
\end{flushleft}

In what follows, we still consider the classical Hamiltonian framework. The following theorem is a generalization of the main result in \cite{GGL}.

\mbox{}

\noindent {\bf Theorem 1.}
{\em Consider a Hamiltonian system defined in $\mathbb{R}^{2m}$ with Hamiltonian given by
\begin{equation}\label{eq2.1-Darboux-Poisson-1}
H(q, p) = \frac{1}{2} \sum_{i=1}^{m} \mu_i \left( \sum_{j=1}^{m} a_{ij} p_j  \right)^2+ V \left(A^T \cdot q \right)
\end{equation}
where $A = (a_{ij})$ is any nonsingular real square matrix of order $m$. If the potential $V(q)$ is a polynomial of degree at least $3$ and the system possesses a proper Darboux polynomial $F(q,p)$, then $H_F(q,p)\, = \, F(q,p)
F(q,-p)$ is a polynomial first integral of the system. Moreover, if $m \geq 2$ and at least two $\mu_i$ are not zero, then
$H_F$ is an additional polynomial first integral.
}

\mbox{}

\noindent{\bf Proof.}
We perform a linear canonical transformation $(q, p)^T = B \cdot (Q, P)^T$ where we shall use the following block diagonal structure
$$
B = \left( \begin{array}{cc} (A^{-1})^T & O_m \\ O_m & A \end{array} \right)
$$
In particular, $B$ is a real symplectic matrix of order $2 m$, that is satisfying by definition the condition $\mathcal{S}_{2m} = B \cdot \mathcal{S}_{2m} \cdot B^T$. After the canonical transformation, the Hamiltonian becomes natural
$$
H^*(Q, P) = \frac{1}{2} \sum_{i=1}^{m} \mu_i P_i^2 + V(Q) \ .
$$
Under the hypothesis of the theorem, the new Hamiltonian system has a proper Darboux polynomial $F^*(Q, P) = F(B \, (Q, P)^T)$. Then, using the results of \cite{GGL} it is found that $H_{F^*}(Q,P)\, = \, F^*(Q, P) F^*(Q,-P)$ is a polynomial first integral of the new Hamiltonian system. Finally, reversing the change of variables, we get that $H_F(q,p)$ is a polynomial first integral of the Hamiltonian system defined by (\ref{eq2.1-Darboux-Poisson-1}). The fact that $H_F$ is additional follows from the hypothesis $m \geq 2$ and the existence of at least two $\mu_i$ not zero, using again \cite{GGL}.
$\:\;\:\; \Box$

\mbox{}

Notice that, the particular case where $A = I_m$ in Theorem 1 is just the one studied in \cite{GGL}. This motivates a more general approach, which is the goal of the next section.

\mbox{}

\begin{flushleft}
{\bf 3. Case II: Diffeomorphic transformation without Casimir decoupling}
\end{flushleft}

As seen in Theorem 1, the use of suitable tranformations leads to the possibility of generalizing previously known results. However, the natural setting for this approach is not the Hamiltonian one. The reason is that canonical transformations are too restrictive. Of course, this difficulty reflects the very special form of Hamilton's equations. However, this perspective is generalized in the Poisson systems context, in which every diffeomorphic transformation is admissible, as indicated in the Introduction. This allows a more general approach to the problem.

\mbox{}

\noindent {\bf Theorem 2.}
{\em Assume that we have a Poisson system
\begin{equation}\label{eq2-Darboux-Poisson}
\dot{x} = \mathcal{J}(x) \cdot \nabla H(x)
\end{equation}
defined in a domain $\Omega \subseteq \mathbb{R}^{2m}$ with Hamiltonian given by
\begin{equation}
H(x) = \frac{1}{2} \sum_{i=1}^{m} \mu_i \Phi_{m+i}^2(x) + V(\Phi_1(x), \ldots ,\Phi_m(x))
\end{equation}
and satisfying the following set of conditions.
\begin{itemize}
\item[(i)] $\Phi(x) = (\Phi_1(x), \ldots ,\Phi_{2m}(x))$ is a $C^{1}$ diffeomorphism in $\Omega$.

\item[(ii)] The structure matrix has the form $\mathcal{J}(x) = M(\Phi(x)) \cdot \mathcal{S}_{2m} \cdot M^T(\Phi(x))$ where $M$ denotes the Jacobian matrix of the transformation $\Phi^{-1}$.

\item[(iii)] The potential $V$ is a polynomial of degree at least 3 in the variables $\Phi_i$ with $i=1, \ldots, m$.

\item[(iv)] The Poisson system admits a proper Darboux invariant hypersurface with respect to $\Phi$ described by the equation $F(x)=0$.
\end{itemize}
Then we have that
$$
H_F(x) := F(\Phi(x)) \, F(\Phi_1(x), \ldots, \Phi_m(x), -\Phi_{m+1}(x), \ldots, -\Phi_{2m}(x))
$$
is a first integral of (\ref{eq2-Darboux-Poisson}). Moreover, if $m \geq 2$ and at least two $\mu_i$ are not zero, then
$H_F(x)$ is an additional first integral of (\ref{eq2-Darboux-Poisson}).
}

\mbox{}

\noindent {\bf Proof.}
We start by noting that system (\ref{eq2-Darboux-Poisson}) is Poisson because matrix $\mathcal{J}(x)$ is by construction a maximal rank structure matrix, namely it is invertible, skew-symmetric and is a solution of the Jacobi partial differential equations. Now the following diffeomorphic transformation is applied:
\[
	q_i = \Phi_i(x) \;\: , \:\;\:\; p_i = \Phi_{m+i}(x) \;\: , \:\;\:\; i = 1, \ldots ,m \ .
\]
This diffeomorphism is, in general, not a canonical transformation from the point of view of classical Hamiltonian systems. However, the transformed system is always a Poisson system and, in fact, in this case it will be Hamiltonian as we shall see. Due to this, the Poisson framework is natural when general transformations have to be considered.

We perform the transformation $x = \Phi^{-1}(q,p)$. The structure matrix of the transformed system will be $\mathcal{S}_{2m}$ due to the relationship:
\begin{equation} \label{bocata}
\mathcal{S}_{2m} = M^{-1}(\Phi(x(q,p))) \cdot \mathcal{J}(x(q,p)) \cdot M^T(\Phi(x(q,p)))
\end{equation}
Accordingly, the Poisson system (\ref{eq2-Darboux-Poisson}) becomes a natural Hamiltonian polynomial system of $m$ degrees of freedom having the Hamiltonian function $H^*(q, p)$ of the form:
\begin{equation}\label{eq1-Darboux-Poisson}
H^*(q,p)= \frac{1}{2} \sum_{i=1}^{m} \mu_i p_i^2 + V(q_1, \ldots ,q_m) \ .
\end{equation}
This Hamiltonian system has the algebraic invariant hypersurface $F(q,p)=0$, where $F(q,p)$ is a proper Darboux polynomial because $F(x)$ is by hypothesis not a first integral of system (\ref{eq2-Darboux-Poisson}). Then it is possible to make use of Theorem 3 of Reference \cite{GGL}. The outcome is that $F(q,p) \, F(q,-p)$ is a polynomial first integral of our Hamiltonian system and, moreover, if $m \geq 2$ and at least two $\mu_i$ are not zero, then it is an additional first integral of such system. Reversing the former change of variables the result is shown.
$\:\;\:\; \Box$

\mbox{}

The results developed in Theorem 2 can be now illustrated by means of some examples.

\mbox{}

\noindent {\bf Example 1.}

\mbox{}

In what follows $m=2$ is chosen, and we define the diffeomorphism $\Phi(x)$ of $\mathbb{R}^4$ whose components $\Phi(x) = (\Phi_1(x), \Phi_{2}(x), \Phi_{3}(x), \Phi_{4}(x))$ are defined by $\Phi_1(x) = x_1$, $\Phi_2(x) = x_2 + x_1 x_2 + x_1 x_3 + x_2 x_3 + x_1 x_2 x_3 + x_1 x_4 + x_3 x_4 + x_1 x_3 x_4$, $\Phi_3(x) = x_3$ and $\Phi_4(x) = x_4 -x_1 x_2 - x_2 x_3 - x_1 x_2 x_3 - x_1 x_4 - x_3 x_4 - x_1 x_3 x_4$. The inverse $\Phi^{-1}(y)$ has components $\Phi_1(y) = y_1$, $\Phi_2(y) = y_2 - y_1 y_2 - y_1 y_3 + y_1^2 y_3 - y_2 y_3 - y_1 y_2 y_3 + y_1 y_3^2 + y_1^2 y_3^2 - y_1 y_4 - y_3 y_4 - y_1 y_3 y_4$, $\Phi_3(y) = y_3$ and $\Phi_4(y) = y_4 + y_1 y_2 - y_1^2 y_3 + y_2 y_3 + y_1 y_2 y_3 - y_1 y_3^2 - y_1^2 y_3^2 + y_1 y_4 + y_3 y_4 + y_1 y_3 y_4$. Then, the skew-symmetric structure matrix $\mathcal{J}(x) = M(\Phi(x)) \cdot \mathcal{S}_{2m} \cdot M^T(\Phi(x))$ of order 4 has the following nontrivial entries:
\begin{eqnarray*}
J_{12} &=& -x_2 + x_1^2 (1 + x_3) - x_4 - x_1 (1 + x_2 - x_3 + x_4) \ , \\
J_{13} &=& 1 \ , \\
J_{14} &=& x_2 - x_1^2 (1 + x_3) + x_4 + x_1 (x_2 - x_3 + x_4) \ , \\
J_{23} &=& (1 + x_1) x_3^2 - x_2 (1 + x_3) + x_3 (-1 + x_1 - x_4) - x_4 \ , \\
J_{24} &=& 1 - x_2 x_3 - x_3 x_4 + x_1 (x_2 + x_4) \ , \\
J_{34} &=& x_3^2 - x_2 (1 + x_3) + x_1 x_3 (1 + x_3) - x_4 - x_3 x_4 \ .
\end{eqnarray*}

\mbox{}

\noindent {\bf Example 2.}

\mbox{}

Consider now natural polynomial Hamiltonian systems (\ref{eq1-Darboux-Poisson}) with two degrees of freedom ($m=2$), $\mu_1 = \mu_2$ and with a potential $V(q_1, q_2)$ of degree 4. The list of integrable systems of this form with homogeneous potential and a first integral of degree at most 4 in momenta is given in \cite{H}. If $V(q_1, q_2) = c q_1^2 + q_2^4$ with $c$ a constant, then an additional first integral is $I(q, p) = p_2^2 + 2 q_2^4$. This first integral comes from the proper Darboux polynomial $F(q, p) = i p_2 + \sqrt{2} q_2^2$ with $i^2 = -1$ and having associated cofactor $K(q, p) = 2 i \sqrt{2} q_2$. The reason is that $I(q, p) = F(q, p) F(q, -p)$.

\mbox{}

\noindent {\bf Example 3.}

\mbox{}

The previous results can be applied also within the framework of classical Hamiltonian theory. For instance, consider a non-natural two-degree of freedom Hamiltonian
$$
H(q, p) = \frac{1}{2} \sum_{i=1}^{2} \mu_i p_i^2 + V(q- \nabla \xi(p)) \ ,
$$
for some smooth scalar function $\xi(p)$. It can be shown that the change of variables $(p, q) \mapsto (P, Q)$ defined by $Q = q - \nabla \xi(p)$, $P=p$ is in fact a canonical transformation. Therefore, in the new variables the resulting Hamiltonian $H^*(Q, P)$ becomes natural
$$
H^*(Q, P) = \frac{1}{2} \sum_{i=1}^{2} \mu_i P_i^2 + V(Q) \ .
$$
In case that $V(Q) \in \mathbb{R}[Q]$, Theorem 2 can be used in order to detect possible additional first integrals.

\mbox{}

\begin{flushleft}
{\bf 4. Case III: Diffeomorphic transformation with Casimir decoupling}
\end{flushleft}

\mbox{}

\noindent {\bf Theorem 3.}
{\em Assume that we have a Poisson system
\begin{equation}\label{eq2.1-Darboux-Poisson}
\dot{x} = \mathcal{J}(x) \cdot \nabla H(x)
\end{equation}
defined in a domain $\Omega \subseteq \mathbb{R}^{2m+s}$ with Hamiltonian given by
$$
H(x) = \frac{1}{2} \sum_{i=1}^{m} \mu_i \Phi_{m+i}^2(x) + V(\Phi_1(x), \ldots ,\Phi_m(x)) + W(\Phi_{2m+1}(x), \ldots ,\Phi_{2m+s}(x))
$$
and satisfying the following set of conditions.
\begin{itemize}
\item[(i)] $\Phi(x) \equiv \Phi_H(x) \oplus \Phi_D(x) = (\Phi_1(x), \ldots ,\Phi_{2m}(x)) \oplus (\Phi_{2m+1}(x), \ldots ,\Phi_{2m+s}(x))$ is a $C^{1}$ diffeomorphism in $\Omega$.

\item[(ii)] The structure matrix has the form $\mathcal{J}(x) = M(\Phi(x)) \cdot \mathcal{S}_{2m,s} \cdot M^T(\Phi(x))$ where matrix $M$ denotes the Jacobian matrix of the transformation $\Phi^{-1}$.

\item[(iii)] The potential $V$ is a polynomial of degree at least 3 in the variables $\Phi_i$ with $i=1, \ldots, m$.

\item[(iv)] The Poisson system admits a proper Darboux invariant hypersurface with respect to $\Phi_H(x)$ described by the equation $F(x)=0$.
\end{itemize}
Then we have that
$$
H_F(x) := F(\Phi_H(x)) \, F(\Phi_1(x), \ldots, \Phi_m(x), -\Phi_{m+1}(x), \ldots, -\Phi_{2m}(x))
$$
is a first integral of (\ref{eq2.1-Darboux-Poisson}). Moreover, if $m \geq 2$ and at least two $\mu_i$ are not zero, then
$H_F(x)$ is a first integral of (\ref{eq2.1-Darboux-Poisson}) additional to both the Hamiltonian $H(x)$ and the complete set of independent Casimir invariants $\Phi_{2m+1}(x), \ldots ,\Phi_{2m+s}(x)$.
}

\mbox{}

\noindent {\bf Proof.}
Notice first that system (\ref{eq2.1-Darboux-Poisson}) is Poisson since matrix $\mathcal{J}(x)$ is by construction (see statement (ii)) a structure matrix. Analogously to the procedure followed in Theorem 2, now the following diffeomorphic transformation is applied:
\begin{eqnarray*}
q_i &=& \Phi_i(x) \;\: , \:\;\:\; p_i = \Phi_{m+i}(x) \;\: , \:\;\:\; i = 1, \ldots ,m \\
z_j &=& \Phi_{2m+j}(x) \;\: , \:\;\:\; j = 1, \ldots ,s .
\end{eqnarray*}
Thus the system is transformed into another Poisson system that is in the Darboux canonical form. In particular, the new Poisson system is given in terms of Hamiltonian
\begin{equation}
\label{calamares}
	\hat{H}(q,p,z) \equiv H^*(q,p)+W(z) = \frac{1}{2} \sum_{i=1}^{m} \mu_i p_i^2 +
	V(q_1, \ldots ,q_m) + W(z_1, \ldots ,z_s)
\end{equation}
and structure matrix
\begin{equation}
\label{pantumaca}
	\hat{\mathcal{J}}(q,p,z) = \mathcal{S}_{2m,s}
\end{equation}
Note that now the variables $z_1, \ldots ,z_s$ are Casimir invariants of this system. Due to this, function $W(z_1, \ldots ,z_s)$ is also a Casimir invariant. To complete the reduction, the Casimir invariants are decoupled, the outcome being an $m$-degree of freedom and natural polynomial Hamiltonian system defined in terms of $H^*(q,p)$ as already defined. By hypothesis, this Hamiltonian system has the algebraic invariant hypersurface $F^*(q,p)=0$, where $F^*(q,p)$ is a proper Darboux polynomial. Therefore use can be made of Theorem 3 of Reference \cite{GGL}. Consequently $H_F^*(q,p)=F^*(q,p) \, F^*(q,-p)$ is a polynomial first integral of the Hamiltonian system defined by $H^*(q,p)$ and, moreover, if $m \geq 2$ and at least two $\mu_i$ are not zero, then it is an additional first integral. The same Darboux polynomial $F^*(q,p)$ and polynomial first integral $H_F^*(q,p)$ are pulled back for the Poisson system (\ref{calamares}-\ref{pantumaca}). In particular, $H_F^*(q,p)$ is obviously independent of $\hat{H}(q,p,z)$, $H^*(q,p)$ and the Casimir invariants $z$. To complete the proof, transformation $\Phi^{-1}$ is applied to system (\ref{calamares}-\ref{pantumaca}).
$\:\;\:\; \Box$

\mbox{}

It is worth noting that in Theorem 3 we retrieve the symplectic formulation of the previous section as a particular case for which $s=0$. The results developed in Theorem 3 are also illustrated in what follows.

\mbox{}

\noindent {\bf Example 4.}

\mbox{}

We now illustrate how Theorem 3 works in the case that diffeomorphism $\Phi$ is a linear separable transformation. Thus, consider the $(2m+s)$-dimensional Poisson system of variables $(q,p,z) \in \mathbb{R}^m \times \mathbb{R}^m \times \mathbb{R}^s$ defined in terms of a Hamiltonian function
\begin{equation}
\label{pechuga}
H(q, p,z) = \frac{1}{2} \sum_{i=1}^{m} \mu_i \left( \sum_{j=1}^{m} c_{ij} p_j  \right)^2+ V \left(B \cdot q \right)+W(D \cdot z)
\end{equation}
where $C=(c_{ij})$, $B$ and $D$ are invertible real matrices of orders $m$, $m$ and $s$, respectively, while the structure matrix is of the form:
\begin{equation}
\label{muslo}
	\mathcal{J} = A^{-1} \cdot \mathcal{S}_{2m,s} \cdot (A^{-1})^T
\end{equation}
where $A=B \oplus C \oplus D$. Assume that this Poisson system has a polynomial potential $V$ of degree at least 3, and moreover it admits an invariant algebraic hypersurface described by the proper Darboux polynomial $F(q,p)$ and being of the form $F(q,p) = F^*(B \cdot q, C \cdot p)$ where $F^*$ is a polynomial in the variables $B \cdot q$ and $C \cdot p$.

The following linear transformation is now applied:
\[
    \left( \begin{array}{c} Q \\ P \\ Z \end{array} \right) = A \cdot \left( \begin{array}{c} q \\ p \\ z \end{array} \right)
\]
The resulting Poisson system is now given by a structure matrix $\hat{\mathcal{J}}$ and Hamiltonian function:
\[
\hat{H}(Q,P,Z) = H^*(Q,P) + W(Z) =  \frac{1}{2} \sum_{i=1}^{m} \mu_i P_i^2+ V(Q)+W(Z)
\]
The $s$ variables $Z$ are now Casimir invariants and can thus be decoupled, the outcome being a classical Hamiltonian system of $m$ degrees of freedom and natural Hamiltonian function $H^*(Q,P)$. Moreover, this Hamiltonian system has the proper Darboux polynomial $F^*(Q,P)$ and consequently the polynomial first integral $H^*_F(Q,P)=F^*(Q,P)F^*(Q,-P)$. As noted, when $m \geq 2$ and at least two $\mu_i$ are not zero, such first integral is additional. Therefore, going back through the performed change of variables, we conclude that the Poisson system defined by (\ref{pechuga}-\ref{muslo}) has the additional polynomial first integral $H_F(q,p)=F^*(B \cdot q,C \cdot p)F^*(B \cdot q,-C \cdot p)$.

\pagebreak

\end{document}